\documentclass[runningheads]{llncs}
\usepackage[T1]{fontenc}
\usepackage[left=105pt, right=105pt,bottom=125pt]{geometry}
	\usepackage{graphicx} 
        \usepackage{pdfcomment}
        \usepackage{calc}
	\usepackage{balance}
	\usepackage{amssymb, amsmath}
	\usepackage{multirow, rotating, wasysym, url}
	\usepackage{subcaption}
 \usepackage[export]{adjustbox}
	\usepackage[ruled,linesnumbered,vlined]{algorithm2e}
	\usepackage{hyperref} 
	\usepackage{color}

\graphicspath{{./DrawingFigures/},{./EvaluateFigures/}} 

	\newcommand{\presec}{\vspace{0.00in}}
	\newcommand{\postsec}{\vspace{0.00in}}
	\newcommand{\presub}{\vspace{0.00in}}
	\newcommand{\postsub}{\vspace{0.00in}}

	\newcommand{\postfig}{\vspace{-0in}}

		\mathchardef\Gamma="0100 \mathchardef\Delta="0101
\mathchardef\Theta="0102 \mathchardef\Lambda="0103
\mathchardef\Xi="0104 \mathchardef\Pi="0105
\mathchardef\Sigma="0106 \mathchardef\Upsilon="0107
\mathchardef\Phi="0108 \mathchardef\Psi="0109
\mathchardef\Omega="010A

\newcommand{\outline}[1]{}

\usepackage{cite}
\usepackage{xspace}
\usepackage{url}
\usepackage{graphicx}
\usepackage{latexsym}
\usepackage{amssymb}
\usepackage{amsfonts}
\usepackage{psfrag}
\usepackage{wrapfig}
\usepackage{comment}
\usepackage{alltt}
\usepackage{color}

\newcommand{\ie}{\emph{i.e.}\xspace}

\newcommand{\Comment}[1]{}

	\newcommand{\bbb}{\noindent\textbf}

	\definecolor{greener}{RGB}{0,166,0}
	\definecolor{reder}{RGB}{255,0,0}
	\definecolor{bluer}{RGB}{0,0,255}
	
\begin{document}

\title{2FA Sketch: Two-Factor Armor Sketch for Accurate and Efficient Heavy Hitter Detection in Data Streams}

\titlerunning{2FA Sketch: ... Accurate and Efficient Heavy Hitter Detection ...}


    
\author{Xilai Liu\inst{1,2} \and
Xinyi Zhang\inst{3} \and
Bingqing Liu\inst{4} \and
Tao Li\inst{5} \and 
Tong Yang\inst{6} \and
Gaogang Xie\inst{3}}
%
\authorrunning{X. Liu et al.}
%
\institute{Institute of Computing Technologies, Chinese Academy of Sciences\and 
University of Chinese Academy of Sciences, China\hfill \email{liuxilai20121013$@$gmail.com}\\
\and
 CNIC, Chinese Academy of Sciences, China \quad 
 \email{\{xyzhang,xie\}$@$cnic.cn}
 \\ 
 \and
 Beijing University of Posts and Telecommunications, China
\hfill\email{bingqingliu9925$@$163.com}\\ \and
 National University of Defense Technology, China \quad \email{taoli\_network@163.com}\\ \and 
 Peking University, China  \quad \email{yangtongemail$@$gmail.com}}

%

\maketitle
\sloppy
	\begin{abstract}

Detecting heavy hitters, which are flows exceeding a specified threshold, is crucial for network measurement, but it faces challenges due to increasing throughput and memory constraints. Existing sketch-based solutions, particularly those using Comparative Counter Voting, have limitations in efficiently identifying heavy hitters. This paper introduces the Two-Factor Armor (2FA) Sketch, a novel data structure designed to enhance heavy hitter detection in data streams. 2FA Sketch implements dual-layer protection through an improved \texttt{Arbitration} strategy for in-bucket competition and a cross-bucket conflict \texttt{Avoidance} hashing scheme. By theoretically deriving an optimal $\lambda$ parameter and redesigning $vote^+_{new}$ as a conflict indicator, it optimizes the Comparative Counter Voting strategy. Experimental results show that 2FA Sketch outperforms the standard Elastic Sketch, reducing error rates by 2.5 to 19.7 times and increasing processing speed by 1.03 times.


\keywords{Heavy Hitter \and Sketch \and Data Stream.}
\end{abstract}

	\presec
\section{Introduction} \postsec

\subsection{Motivation} \postsec

Network measurement is a critical aspect of networking, providing essential data for applications such as load balancing, quality of service (QoS) management, network caching, congestion control, anomaly detection, and intrusion prevention \cite{mumon,shroff2003measurement,bolt2023,UnivMon,jaqen2021,Kun}. 
Within this domain, identifying heavy hitters --- flows whose sizes exceed a specified threshold --- is particularly crucial as these flows often have a disproportionate impact on network performance and security \cite{zhang2010identifying,UnivMon}. These flows are typically identified by a combination of five tuples: source IP/port, destination IP/port, and protocol.

However, heavy hitter detection faces significant challenges in modern networks. 
Detection systems must track millions of flows in real-time, updating counters for each packet arrival. 
The exponential growth in network throughput has outpaced advancements in CPU processing capabilities, while the memory wall --- the growing disparity between processor and memory speeds --- further exacerbates these limitations \cite{memorywall}. 
These factors collectively render traditional measurement approaches increasingly inadequate for contemporary network.

%

In recent years, sketch-based solutions have gained significant attention for addressing resource constraints by offering approximate methods balancing accuracy and efficiency \cite{elastic,UnivMon,chainsketch, sketchvisor,opensketch,flowradar,dai2016noisy,dai2018finding,daipersisitem16,lidan,ldsketch}. These approaches leverage SRAM's capabilities and simple computational operations to meet latency requirements for real-time processing, utilizing CPU power. They divide memory into key-value pairs to record heavy hitter information. However, SRAM's limited size (generally a few megabytes) \cite{kim2018generic} constrains the flows simultaneously monitored, challenging comprehensive capture of all heavy hitters.

Various sketch techniques aim to approach this goal by prioritizing heavy hitters during the allocation of limited memory resources, exploiting frequency disparities between heavy hitters and small flows \cite{heavykeeper, elastic, uss, cocosketch, chainsketch, lisitan}. These techniques primarily rely on arbitration strategies during flow competition, which can be categorized into three main approaches: Exponential Weighted Decaying, Inverse Frequency Replacement, and Comparative Counter Voting (elaborated in Section \ref{sec:related}). 
While recent advancements have optimized the first two strategies for more accurate heavy hitter detection, Comparative Counter Voting, as implemented in Elastic Sketch \cite{elastic}, has received less attention. 
Unlike probabilistic approaches, this deterministic arbitration strategy offers significant potential for enhancing heavy hitter detection, particularly in scenarios requiring predictability and consistency.

Elastic Sketch employs Comparative Counter Voting primarily to bifurcate flows into two structures: a Light Part with an 8-bit counter CM Sketch \cite{cmsketch} for small flows, and a Heavy Part for larger flows, facilitating heavy hitter detection.
This bifurcation enhances generality, but allocating memory to small flow data in the Light Part potentially constrains key-value pair storage for heavy hitters in the Heavy part, suggesting room for optimization. 
However, naively discarding the Light part would forfeit filtering advantages conferred by this structural separation, resulting in diminished generality without necessarily improving heavy hitter detection efficacy.

\presec
\subsection{Our Solution} \postsec

We introduce the \textbf{Two-Factor Armor (2FA) Sketch}, a novel data structure providing dual-layer protection for enhanced heavy hitter detection, addressing key limitations in existing Comparative Counter Voting approaches.

Comparative Counter Voting, which augments positive votes for matching flow keys and negative votes for mismatches, faces two critical challenges in its current implementation: (1) The empirical values of key parameters ($\lambda$ and $vote^+_{new}$) are inherently linked to the Light Part in Elastic Sketch, making its simple removal ineffective. (2) The design lacks a mechanism to resolve competition among heavy hitters within a bucket, occasionally resulting in the loss of some heavy hitter records.

Our proposed solution addresses these challenges by theoretically deriving an optimal $\lambda$ that preserves the flow order relationship, and redesigning $vote^+_{new}$ to function as an indicator of intense heavy hitter competition within a bucket. Upon detecting such competition, a rehashing mechanism is triggered for conflict avoidance, thereby preventing the omission of heavy hitters. 

The main contributions of this paper are as follows:
\begin{itemize}
    \item An improved \texttt{Arbitration} strategy for in-bucket competition, serving as the first armor layer. This Arbitration strategy, an improved version of Comparative Counter Voting, is applied to flows after the Light Part has been discarded to allocate more space for heavy hitters.
    \item A cross-bucket conflict \texttt{Avoidance} hashing scheme, acting as the second armor layer. This method complements the Arbitration strategy by addressing recall rate degradation caused by excessive flow congestion within individual buckets.
    \item Our experimental results demonstrate that 2FA Sketch significantly outperforms the standard Elastic Sketch, reducing the error rate by 2.5--19.7 times and increasing processing speed by 1.03 times. All source codes are publicly available on GitHub \cite{opensource} to facilitate reproducibility and further research.
\end{itemize}

	\presec
\section{Background and Related Work}\label{sec:related}
\postsec

In this section, we begin with essential definitions of heavy hitters. 
We then briefly introduce the common sketch data structure.
Finally, we focus our discussion on the distinguishing flow competing strategies within these sketches.

\presub
\subsection{Problem Definition}

Data streams are defined as continuous sequences of packets, each represented by a tuple $(f, v_f)$ and processed exactly once. The flow key $f$ comprises any combination of the five-tuple elements: source IP, destination IP, source port, destination port, and protocol number. The value $v_f$ represents either a unit count $1$ or the packet size.

A flow $f$ is classified as a heavy hitter if its accumulated frequency $V_f$ over the data stream exceeds the threshold $\Theta = \theta \cdot N$, where $N$ denotes the total number of packets in the data stream and $\theta$  a predefined fraction, e.g, $0.01$.

\presub
\subsection{Common Sketch Data Structure}\label{sec:common_sketch_structure}

\begin{figure}[htbp]
	\centering
	\includegraphics[width=0.9\textwidth]{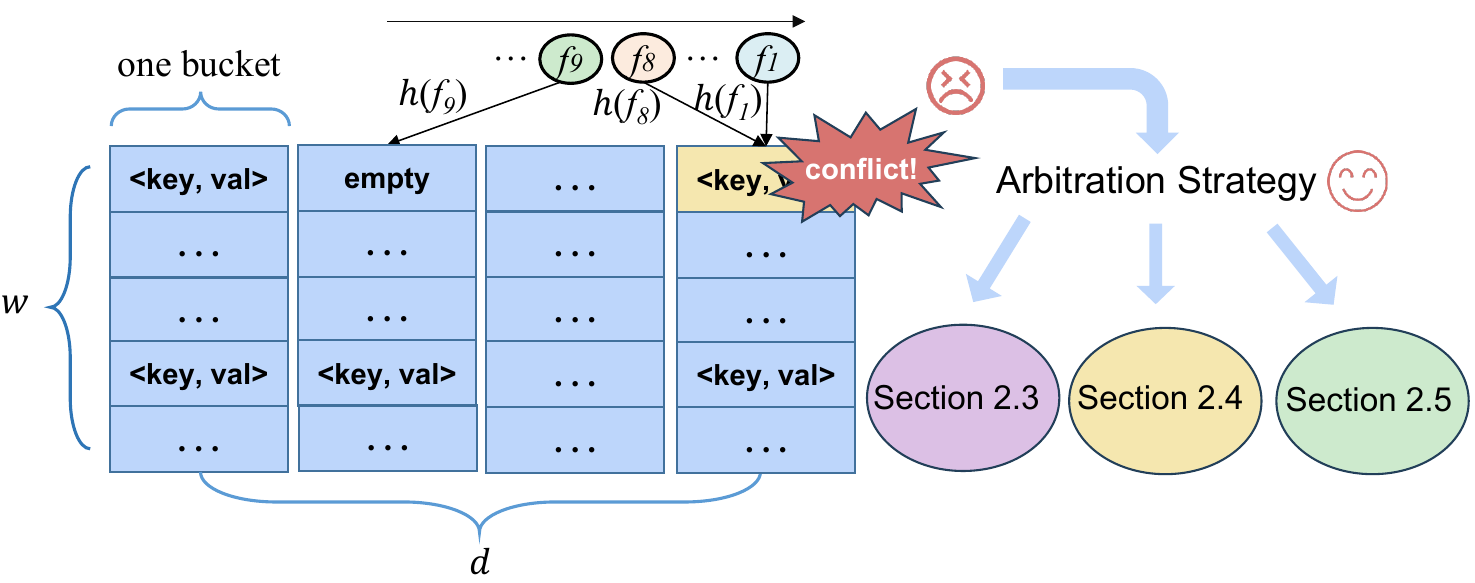}
	\caption{Common Sketch Data Structure.}
	\label{common_sketch_structure}
\end{figure}

The common data structure of sketches for heavy hitter detection comprises a matrix of size $w \times d$, where $d$ denotes the number of buckets, each containing $w$ cells (as shown in Figure \ref{common_sketch_structure})\footnote{Here, structures containing key-value pairs are generally referred to as counter-based methods in previous research, while those containing only values are termed sketch-based methods. For simplicity, we refer to both as "sketches" in this paper. The latter type of methods is discussed in Section \ref{OtherSketch}.}.  This configuration aligns with set-associative cache memory structures. When $w = 1$, the structure is a hash table \cite{hashtable} with elevated collision probability, while $d = 1$ corresponds to the SpaceSaving \cite{SpaceSaving} algorithm, necessitating Stream Summary technique for expedited cell location. These parameters primarily influence processing speed.

Each cell typically stores two fields for a flow: the flow key and its corresponding value. Packet arrival triggers mapping to a hashed bucket. Insertion is straightforward when the bucket contains the flow key or has empty cells. Conflicts arise when the hashed bucket is full and does not contain the flow key. To address this, various methods identify the cell with the minimum value $C$ within the bucket, minimizing conflict. The incoming flow key then undergoes arbitration with the existing flow to determine retention and resulting value. Exploiting the frequency disparity between heavy hitter and small flows, research has yielded three main arbitration strategies: Exponential Weighted Decaying \cite{heavykeeper, heavyguardian}, Inverse Frequency Replacement \cite{drap,uss,cocosketch, chainsketch}, and Comparative Counter Voting \cite{elastic}. The configuration of these strategies predominantly affects processing accuracy.

\presub
\subsection{Exponential Weighted Decaying}
This arbitration strategy was initially introduced in the HeavyGuardian paper \cite{heavyguardian}. For the cell with the minimum value $C$, the count field is decayed by $1$ with probability $P = b^{-C}$, where $b$ is a predefined constant (e.g., $b = 1.08$). Post-decay, if the count field reaches $0$, the new flow $f$ replaces the flow key field and resets the count to $1$; otherwise, $f$ is either discarded or inserted into an alternative structure. This approach amplifies the value differentials between conflicting flows by utilizing the value as an exponent. The HeavyKeeper \cite{heavykeeper} algorithm also incorporated this concept. Subsequently, ActiveKeeper \cite{activekeeper} and HeavyTracker \cite{heavytracker} have further refined the probability expression, building upon HeavyKeeper to enhance accuracy. This strategy is especially suitable to the top-k elephant flow detection task.

\presub
\subsection{Inverse Frequent Replacement}
This approach probabilizes replacement while maintaining deterministic counting, in contrast to the previous method that probabilized counting and determinized replacement. For the cell with the minimum value $C$, counting invariably occurs upon the arrival of a new flow, incrementing by $1$. The key field has a $\frac{1}{C + 1}$ probability of being replaced by the new flow and a $\frac{C}{C + 1}$ probability of remaining unchanged. This method amplifies the tracking differences between flows by utilizing the value in the denominator as the replacement probability. Several algorithms, including USS, CocoSketch, RAP, and ChainSketch \cite{uss, cocosketch,drap, chainsketch}, have implemented this design. These studies demonstrate that under this replacement probability, the expected count of a flow remains unbiased.

\presub
\subsection{Comparative Counter Voting}
In this strategy, each bucket comprises multiple key-value cells (with values serving as $vote^+$) and one final cell without a key, which records negative votes ($vote^-$), collectively forming the Heavy Part in Elastic Sketch. Upon conflict in the minimum-value cell, $vote^-$ increments while $vote^+$ remains static. Replacement occurs when $\frac{vote^-}{vote^+}$ exceeds a preset threshold $\lambda$ (e.g., $4$). During replacement in the Elastic Sketch, $vote^+_{new}$ resets to $1$, $vote^-$ to $0$, and the flow key updates to the new flow ($f$). The original flow's count is then evicted to the Light Part, a CM Sketch. Thus, large flow counts combine the cell's value and potential CM Sketch counts. This approach, utilizing vote comparison between different count values, ensures deterministic replacement and counting mechanisms.

\presub
\subsection{Other Sketches for Heavy Hitters}\label{OtherSketch}
\postsub

Various methods beyond key-value pair-based sketches are employed to identify heavy hitters. Algorithms that eschew key storage, such as CM\cite{cmsketch}, CU\cite{cusketch}, Count\cite{csketch}, and CSM\cite{csmsketch} sketches, often utilize an auxiliary min-heap for heavy hitter maintenance. Among these, the CU sketch achieves the highest accuracy, while the CSM sketch provides optimal processing speed.

	\presec
\section{The 2FA Sketch}
\postsec

In this section, we present the details of 2FA Sketch, including its data structure and operations, design innovations and examples.

\presub
\subsection{Algorithm Overview}
\postsub

\begin{figure}[htbp]
	\centering
	\includegraphics[width=\textwidth]{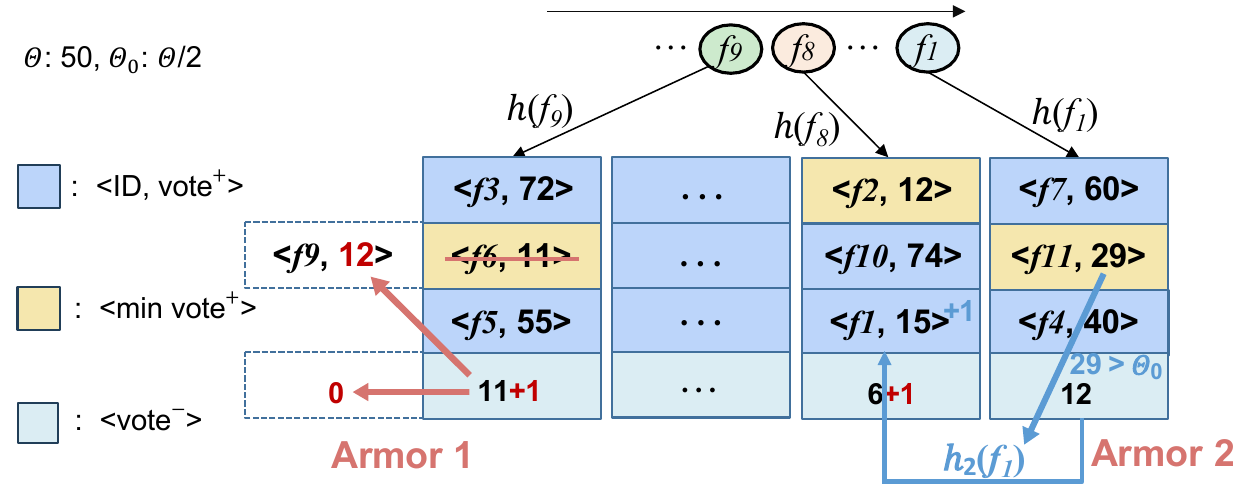}
	\caption{An example for the 2FA Sketch.}
	\label{draw:2FA_Sketch}
\end{figure}

\bbb{Data Structure (Fig. \ref{draw:2FA_Sketch}):} 
The data structure of 2FA Sketch aligns with the structure outlined in Section \ref{sec:common_sketch_structure}, retaining only the Heavy Part of Elastic Sketch. It comprises a bucket array $B$ of size $d$, determined by the available memory and required accuracy constraints.
Each bucket contains $w$ cells: one cell records the negative votes ($vote^-$) for the bucket, while the remaining cells store two fields per flow: the flow key and positive votes ($vote^+$, equivalent to the value field).

\bbb{Insertion (Fig. \ref{draw:2FA_Sketch}):} 
When inserting an incoming packet with flow key $f$, we calculate the hash function, check the hashed bucket $B[h(f)]$ and do the corresponding operations. The pseudo-code without SIMD (Single Instruction Multiple Data) speedup is shown in algorithm \ref{insertion_algo}. After locating the hashed bucket $B[h(f)]$:

\noindent 1) If $f$ matches a flow key within any cell of the bucket, only the corresponding positive votes ($vote^+$) are incremented.

\noindent 2) If  $f$ is not in the bucket, there are the following two cases:

Case 1: There are empty cells in the bucket. We just insert  $f$ with $vote^+$ of 1 into the first empty cell;

Case 2: When no empty cell exists in the bucket, we first increment $B[h(f)].vote^-$ by 1 (hereafter, $B[h(f)]$ is omitted for brevity). We then locate the cell $[min_{index}]$ with the minimum positive votes in the bucket.

\quad 1) If $[min_{index}].vote^+$ exceeds $\Theta_0$ (a threshold related to $\Theta$, e.g., $\Theta/2$), we rehash to an alternative backup bucket $B[h_2(f)]$ and restart the insertion process (only once). 

\quad 2) Otherwise, we compare $vote^-$ with $[min_{index}].vote^+$. If $\frac{vote^-}{[min_{index}].vote^+} > \lambda=1$, $[min_{index}].key$ is replaced by $f$ and $[min_{index}].vote^+$ (equivalent to $vote^+_{new}$) replaced by $vote^-$; otherwise, $f$ is discarded.



\SetAlgoSkip{}
\begin{algorithm}[htbp]
\caption{Insertion Procedure}\label{insertion_algo}
\DontPrintSemicolon  

\KwIn{an incoming flow $f$, threshold $\Theta$}
$FP =$ hash($f$), $pos =FP$ $\%$ $d$, $min_{val} = \infty$, $min_{index} = 0$, $\Theta_0 = \Theta / 2$, $\lambda = 1$\;
\For{$i \gets 1$ \KwTo $w-1$}{
    \If{$FP == B[pos].key[i]$}{
        $B[pos].vote[i] += 1$ \tcp*{$vote[i]$ represents $vote^+$ in loop}
        return\;
    }
    \If{$B[pos].val[i] < min_{val}$}{
        $min_{index} = i$\;
        $min_{val} = B[pos].vote[i]$
    }
}

\If(\tcp*[f]{Case 1}){$min_{val} == 0$}{ 
    $B[pos].key[min_{index}] = f$,
    $B[pos].vote[min_{index}] = 1$\;
    return\;
}
\If(\tcp*[f]{Case 2.1, $vote^+_{new}$'s judgment}){$min_{val} >= \Theta_0$}{ 
    hash $= h_2$,
    goto line 1\;
}

$guard_{val} = B[pos].vote[0] + 1$
\tcp*{$vote[0]$ represents $vote^-$}
\If(\tcp*[f]{Case 2.2}){$\frac{guard_{val}}{min_{val}} >= \lambda$}{
    $B[pos].vote[0] = guard_{val}$\;
}\Else{
    
    $B[pos].key[min_{index}] = f$, $B[pos].vote[0] = 0$\;
    $B[pos].vote[min_{index}] = guard_{val}$\tcp*{$vote^+_{new}$'s update}
}

\end{algorithm}
\bbb{Query:} 
For a single flow $f$, querying whether it is a heavy hitter follows a process similar to insertion. The algorithm locates the buckets $B[h(f)]$ and $B[h_2(f)]$ using hash functions. If $f$ matches a flow key within any cell of these buckets, we obtain the corresponding $vote^+$ value. The flow is reported as a heavy hitter if this value exceeds the threshold $\Theta$. Notably, the information for f is stored in at most one of the buckets $B[h(f)]$ or $B[h_2(f)]$.

To identify all heavy hitters, the algorithm traverses all cells within each bucket, querying the size of each flow. Any flow with a size surpassing the predetermined threshold is classified and reported as a heavy hitter.

\presub
\subsection{Design Innovations}
\postsub

\bbb{Armor 1 (Fig. \ref{draw:2FA_Sketch}):}
Our first Armor layer is an improved \texttt{Arbitration} strategy for in-bucket competition, focusing on two key parameters. The assignment of $vote^+_{new}$ determines the initial frequency count for a new flow during replacement. We have $vote^+_{new} \leq vote^-$, with equality possible only when two flows compete within a single cell. Therefore, we set $vote^+_{new}$ to $vote^-$ instead of $0$ to ensure that any flow recorded in the bucket is not underestimated.

The parameter $\lambda$ governs flow eviction criteria. To preserve flow order relationships, $\lambda$ must not be less than the smallest bucket, ensuring only keys with higher frequency counts in the bucket are recorded (i.e., $\lambda \geq 1$). We set $\lambda$ to the boundary value of 1 as it allows for quick incorporation of potential heavy hitters when $vote^-$ grows rapidly to $[min_{index}].vote^+$, indicating the presence of keys with large frequencies that don't match multiple cells. Elastic's more lenient judgment on this point is due to its ability to count potentially evicted large flows in its Light Part, making it less sensitive to $\lambda$ variations at the cost of relatively larger errors for heavy hitters.

\bbb{Armor 2 (Fig. \ref{draw:2FA_Sketch}):}
Our second Armor layer implements a cross-bucket conflict \texttt{Avoidance} hashing scheme. This approach is founded on two key observations:
\begin{enumerate}
    \item When memory is constrained, hash imbalances and limited bucket numbers can lead to disproportionate flow mappings, resulting in some buckets exceeding their cell capacity while others remain underutilized. This issue cannot be resolved through the first Armor layer's arbitration strategy alone.
    \item The $vote^+_{new}$ value from the arbitration strategy allows us to assess the degree of conflict within a bucket. Throughout the insertion process, $vote^+_{new}$ does not decrease in the replacements, causing the corresponding $[min_{index}].vote^+$ to consistently increase. This metric effectively gauges the intensity of multi-cell conflicts within a bucket.
\end{enumerate}

We recommend the threshold $\Theta_0$ between $0.4\Theta\sim0.8\Theta$. When $[min_{index}].vote^+$ exceeds this threshold, we employ an alternative hash function to redirect to a backup bucket, thereby achieving conflict avoidance.

\bbb{Examples (Fig. \ref{draw:2FA_Sketch}):} 
We use three examples to show the differences between the standard Elastic and our 2FA Sketch in terms of the insertion operation.
Given an incoming packet with flow key $f_1$, the hash function $h(f_1)$ is calculated and the hashed bucket is found. Since flow $f_1$ is not stored in the bucket and  $[min_{index}].vote^+ > \Theta_0$ ($\Theta_0$ sets to $\frac{\Theta}{2}$ here), we rehash to another bucket. At this time, it matches one cell luckily and the corresponding $vote^+$ increases by $1$. 

For another incoming packet with flow key $f_8$, we calculate the hash function and find the hashed bucket. Since $f_8$ is not stored in the bucket, $vote^-(=6)$ is incremented by one. We then find the flow with the smallest size, $f_1$, and find that $vote^- $ is not larger than ${vote^+(=11)}$ of $f_1$. Therefore, $f_8$ is discarded and nothing else is changed.

For another flow $f_9 $, we calculate the hash function and find the hashed bucket. We find flow $f_6$, the flow with the smallest value in the bucket. Because $vote^-(=12 = 11 + 1)$ is larger than ${vote^+(=11)}$ of $f_6$,  $f_6$ is replaced by $f_9$, and the size of $f_9$ is set to $vote^-$ (12, \ie, $vote^+$ of $f_6$ plus 1). Flow $f_6$ is discarded and $vote^-$ is set to zero.

	\presec
\section{Experimental Results}
\postsec

We present experimental results comparing 2FA Sketch with state-of-the-art heavy hitter detection algorithms. Section~\ref{Exp:setup} describes the experimental setup, followed by performance and parameter evaluations in Section~\ref{Exp:eva} and Section~\ref{Exp:parameter}.

\presec
\subsection{Experimental Setup}\label{Exp:setup}
\postsec

\bbb{Datasets}: We have employed two datasets for our study.

\bbb{CAIDA Dataset}:
We use the public traffic trace datasets collected in Equinix-Chicago monitor from CAIDA\cite{caida}. In our experiments, each packet is identified by its source IP address (4 bytes). There are about 25 millions packets in total.


\bbb{Synthetic Dataset}: We generate synthetic datasets with 30 millions packets each, following the \textbf{Zipf} distribution \( p(x) = \frac{x^{-\alpha}}{\zeta(\alpha)} \), with varying \( \alpha \) values ranging from $0.4$ to $1.2$ with a step of 0.2. Zipf’s law, an empirical observation, describes the distribution in phenomena such as the frequency of web page visits. The different \( \alpha \) values represent varying levels of skewness, allowing us to simulate data streams with diverse characteristics.

\presec 
\bbb{Implementation}:
We compare six heavy hitter detection approaches: SpaceSaving (SS)\cite{SpaceSaving}, Count/CM Sketch\cite{cmsketch,csketch} with a min-heap (CountHeap/CMHeap), Elastic Sketch\cite{elastic}, Chain Sketch (Chain) \cite{chainsketch} and 2FA Sketch. All algorithms are implemented in C++ with 100KB default memory size and 0.01\% heavy hitter threshold in every experiment. 
For Elastic and Chain Sketch, we use the parameters in the open-sourced code.
For both Elastic and 2FA Sketch, we store 7 flows and one $vote^-$ in each bucket of the Heavy Part. Elastic uses a 3:1 Heavy to Light Part ratio. The variant of 2FA Sketch employing solely Armor 1 is designated as 1FA. In the implementation of Armor 2, the ratio $\frac{\Theta_0}{\Theta}$ is configured to 0.5.  For CountHeap / CMHeap, we use 3 hash functions for the sketch and set the heap capacity to 4096 nodes. 

\presec
\bbb{Computation Platform}: we conducted all the experiments on a machine with one 4-core processor (4 threads, AMD EPYC 7662@2.0GHz) and 15.4 GB DRAM memory. To accelerate the processing speed, we use SIMD (Single Instruction Multiple Data) instructions for both Elastic and 2FA Sketch. With the AVX2 instruction set, we can find the minimum counter and its index among 8 counters in a single comparison instruction. Also, we can compare 8 32-bit integers with another set of 8 32-bit integers in a single instruction.

\presec
\bbb{Metrics}:
We use the following seven metrics to evaluate the performance of compared algorithms.
\postsec

\bbb{1) AAE (Average Absolute Error):} $\frac{1}{|\Phi|}\sum_{e_i \in \Phi}|f_i-\hat{f_i}|$, where $\Phi$ is the query set of heavy flows and $f_i$ and $\hat{f_i}$ are the actual and estimated flow sizes of flow $f_i$, respectively.

\bbb{2) ARE (Average Relative Error):} $\frac{1}{|\Phi|}\sum_{e_i \in \Phi}\frac{|f_i-\hat{f_i}|}{f_i}$, where $\Phi$, $f_i$, and $\hat{f_i}$ are defined above. 

\bbb{3) PR (Precision Rate):} Ratio of the number of correctly reported flows to the number of reported flows.

\bbb{4) RR (Recall Rate):} Ratio of the number of correctly reported flows to the number of true flows.

\bbb{5)}\ $\mathbf{F_1}$\ \bbb{score:} $\frac{2\times PR\times RR}{PR+RR}$, where PR and RR are defined above.

\bbb{6) Rehashing Ratio:} The proportion of non-matching packets that are rehashed to alternative buckets. 



\bbb{7) Throughput:} million packets per second (Mpps). We use it to evaluate the processing speed of different sketches. The experiments on processing speed are repeated 100 times to minimize accidental errors.

\presec
\subsection{Evaluation on Accuracy and Processing Speed}\label{Exp:eva}
\postsec

\begin{figure*}[hbtp]
\centering
\begin{subfigure}[t]{0.24\textwidth}
\centering
\includegraphics[width=\linewidth, height = 63pt]{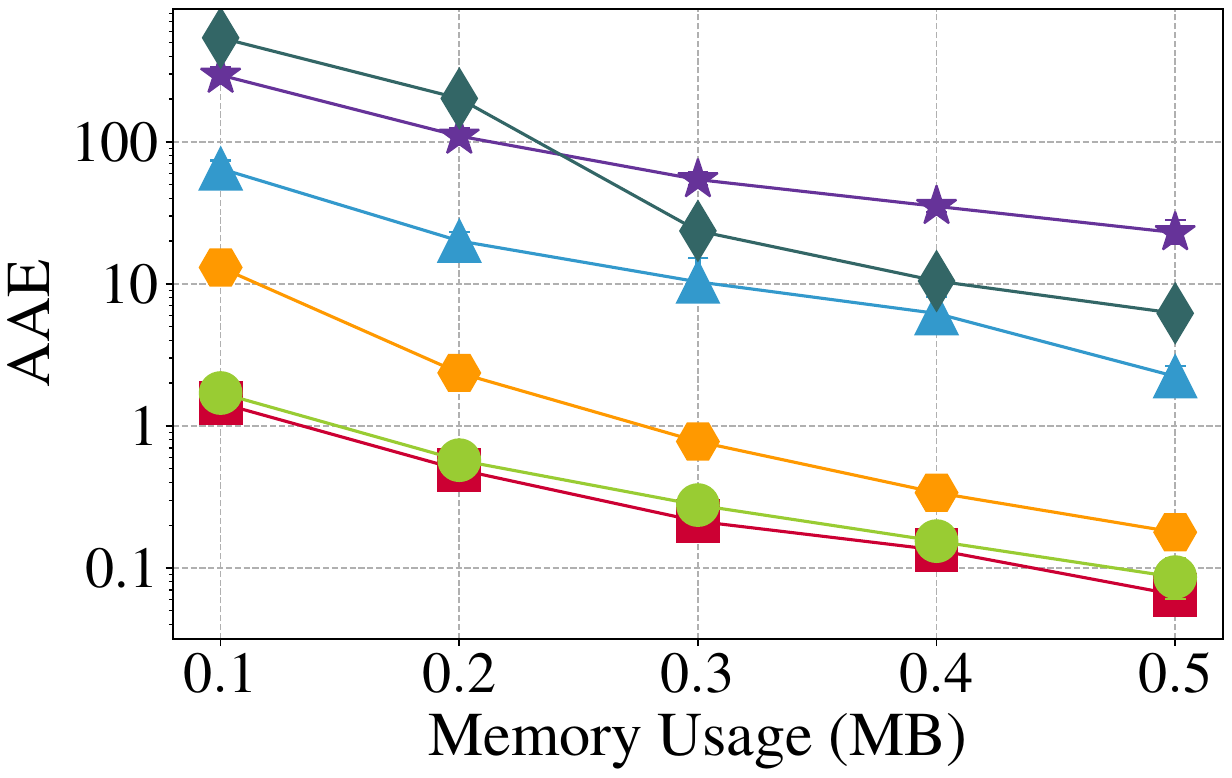}
\caption{AAE}\label{EXP:AAE}
\end{subfigure}\hfill
\begin{subfigure}[t]{0.24\textwidth}
\centering
\includegraphics[width=\linewidth, height = 63pt]{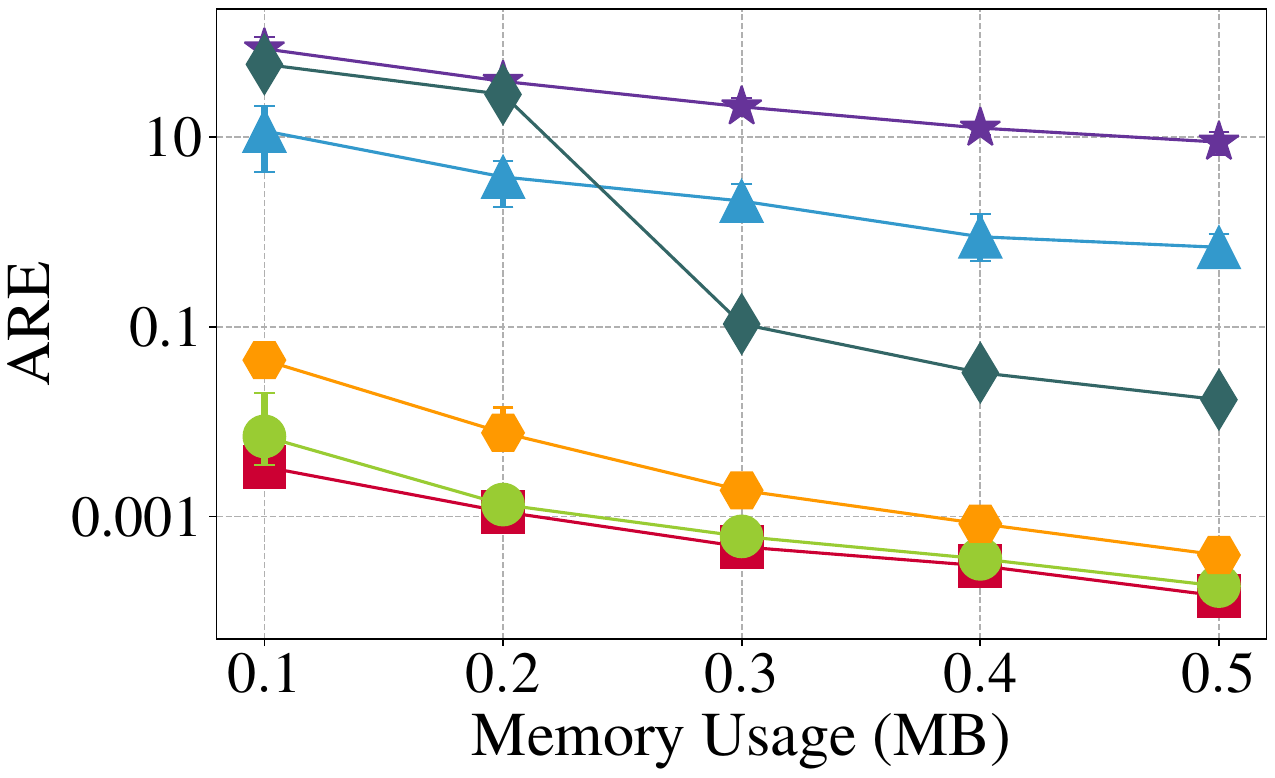}
\caption{ARE} \label{EXP:ARE}
\end{subfigure}\hfill
\begin{subfigure}[t]{0.24\textwidth}
\centering
\includegraphics[width=\linewidth, height = 63pt]{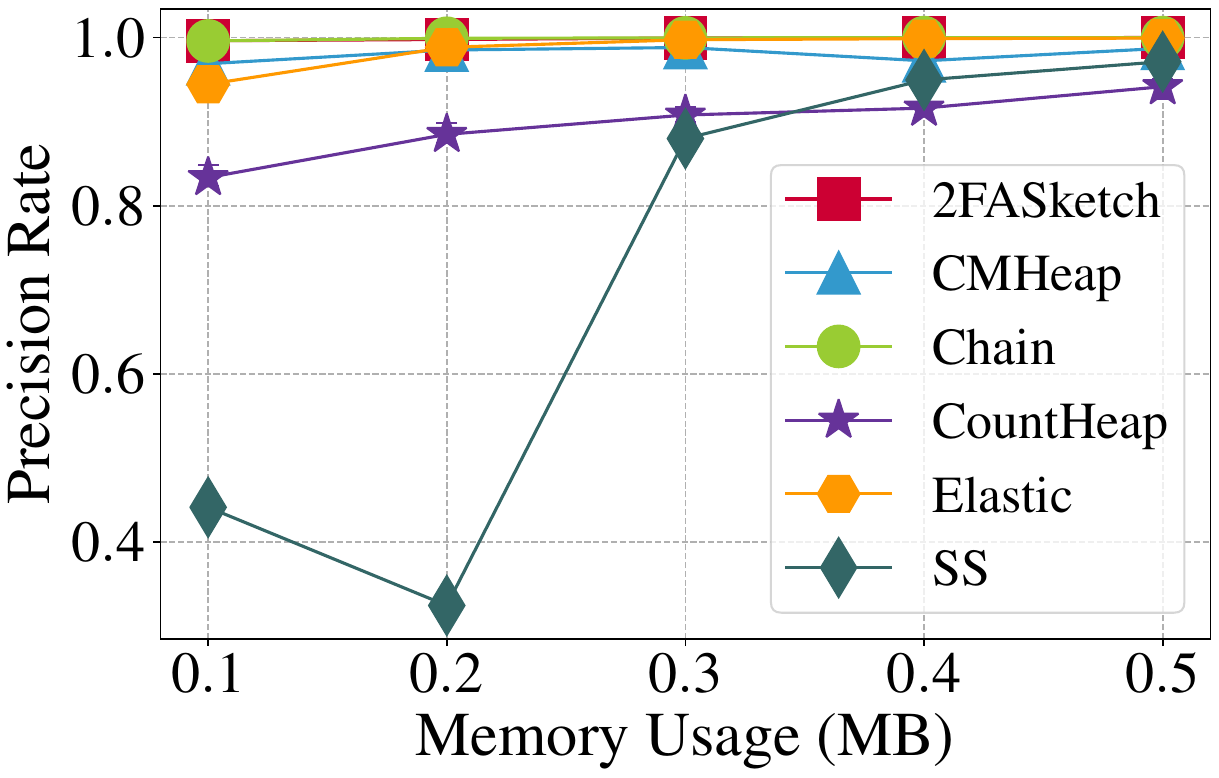}
\caption{precision rate}\label{EXP:precision}
\end{subfigure}\hfill
\begin{subfigure}[t]{0.24\textwidth}
\centering
\includegraphics[width=\linewidth, height = 63pt]{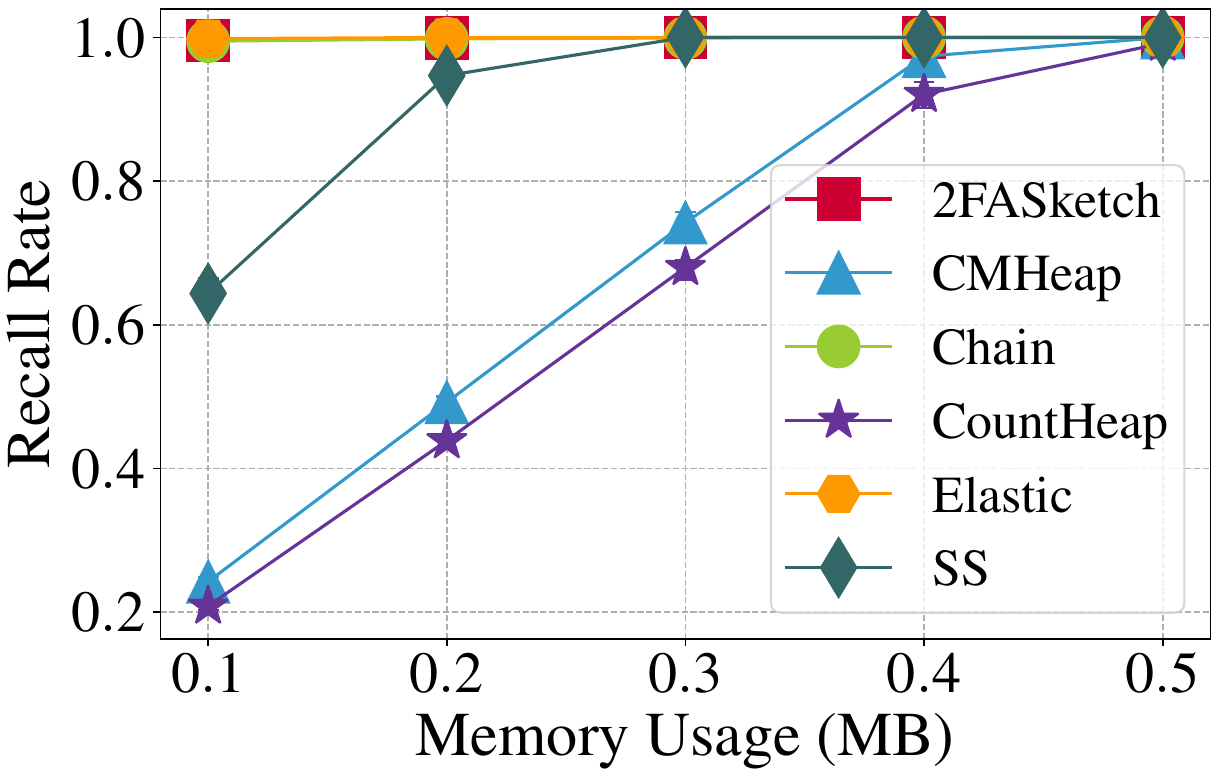}
\caption{recall rate}\label{EXP:recall}
\end{subfigure}
   \caption{Heavy Hitter Detection Accuracy Comparison: CAIDA Dataset (\ref{EXP:AAE} and \ref{EXP:ARE} share the same legend as  the others)}
    \label{fig:accracy one}
    \vspace{-0.3in}
\end{figure*}

\bbb{Accuracy evaluation on CAIDA (Figure~\ref{EXP:AAE}-\ref{EXP:recall}):} With memory sizes of 100KB-500KB, 2FA Sketch outperforms Chain and Elastic in accuracy. 2FA Sketch achieves $7.3\sim8.1$ times lower AAE and $5.7\sim7.5$ times lower ARE than Elastic on CAIDA. 2FA Sketch, Elastic, and Chainsketch maintain nearly $100\%$ PR and RR even at 100KB memory. The other methods, requiring additional memory for min-heap or stream summary structures, perform poorly here but are more suited for top-k elephant flow detection. Subsequently, we evaluate with further reduced memory for 2FA Sketch, Elastic and Chainsketch. 

\newlength{\imageheightA}

\begin{figure*}[hbtp]
\centering
\begin{subfigure}[t]{0.24\textwidth}
\centering
\includegraphics[width=\linewidth, height = 63pt]{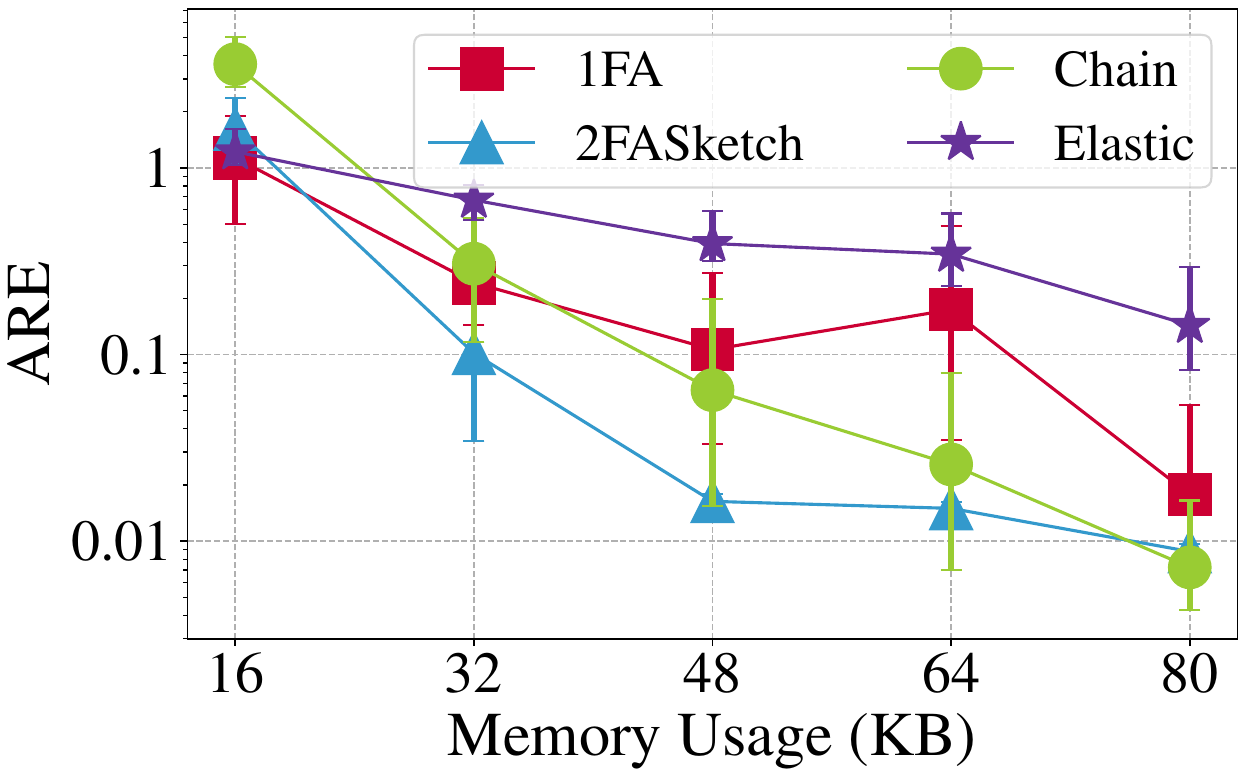}
\caption{ARE}\label{EXP:SM_ARE}
\end{subfigure}\hfill
\begin{subfigure}[t]{0.24\textwidth}
\centering
\includegraphics[width=\linewidth, height = 63pt]{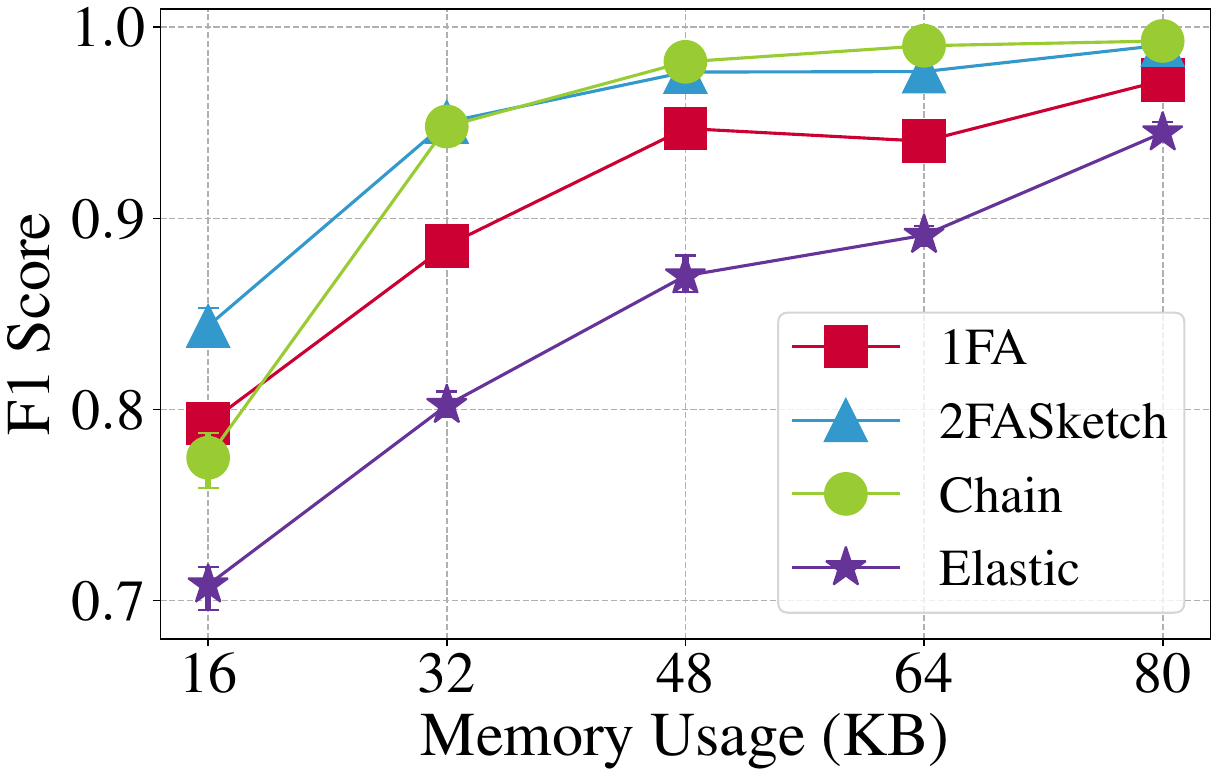}
\caption{F1 Score} \label{EXP:SM_FS}
\end{subfigure}\hfill
\begin{subfigure}[t]{0.24\textwidth}
\centering
\includegraphics[width=\linewidth, height = 63pt]{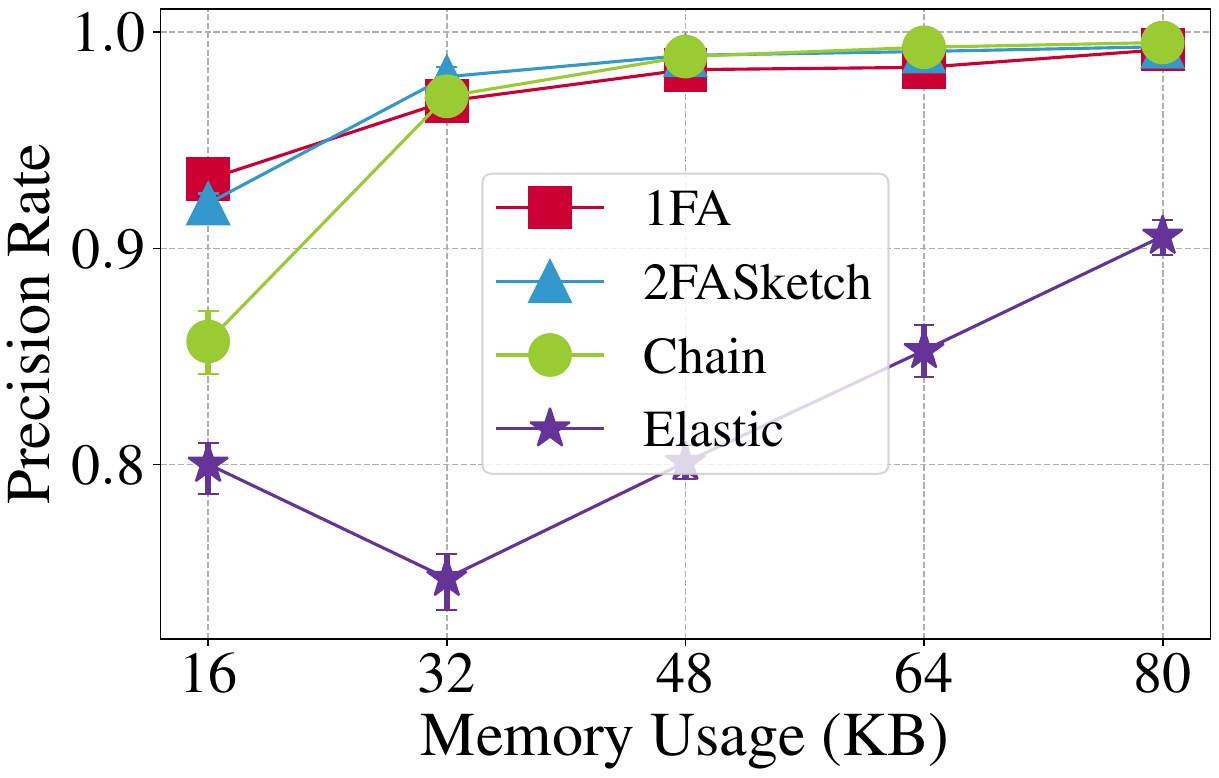}
\caption{precision rate}\label{EXP:SM_precision}
\end{subfigure}\hfill
\begin{subfigure}[t]{0.24\textwidth}
\centering
\includegraphics[width=\linewidth, height = 63pt]{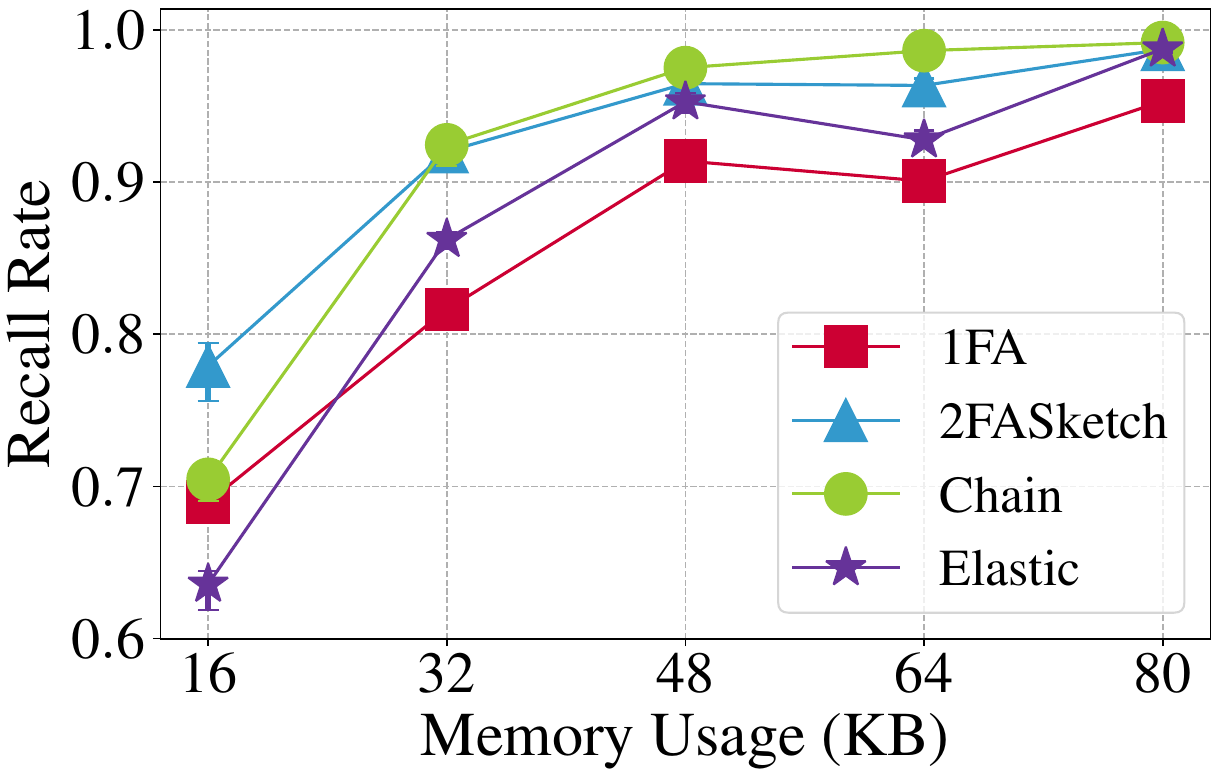}
\caption{recall rate}\label{EXP:SM_recall}
\end{subfigure}
   \caption{Heavy Hitter Detection Accuracy Comparison: CAIDA Dataset Under Memory Constraints}
    \label{fig:accracy_small_memory}
    \vspace{-0.15in}
\end{figure*}

\bbb{Accuracy evaluation on CAIDA under memory constraints (Figure~\ref{EXP:SM_ARE}-\ref{EXP:SM_recall}):} The 2FA Sketch outperforms other methods across all metrics, even with limited memory sizes ranging from 16KB to 80KB. Notably, it achieves an F1 score of approximately $85\%$ with only 16KB of memory. This performance can be attributed to the significant improvement (over $10\%$) of Armor 2 over Armor 1, which was achieved without compromising the ARE and PR for detected flows.

\begin{figure}[htbp]
    \centering
    \includegraphics[width=0.5\textwidth]{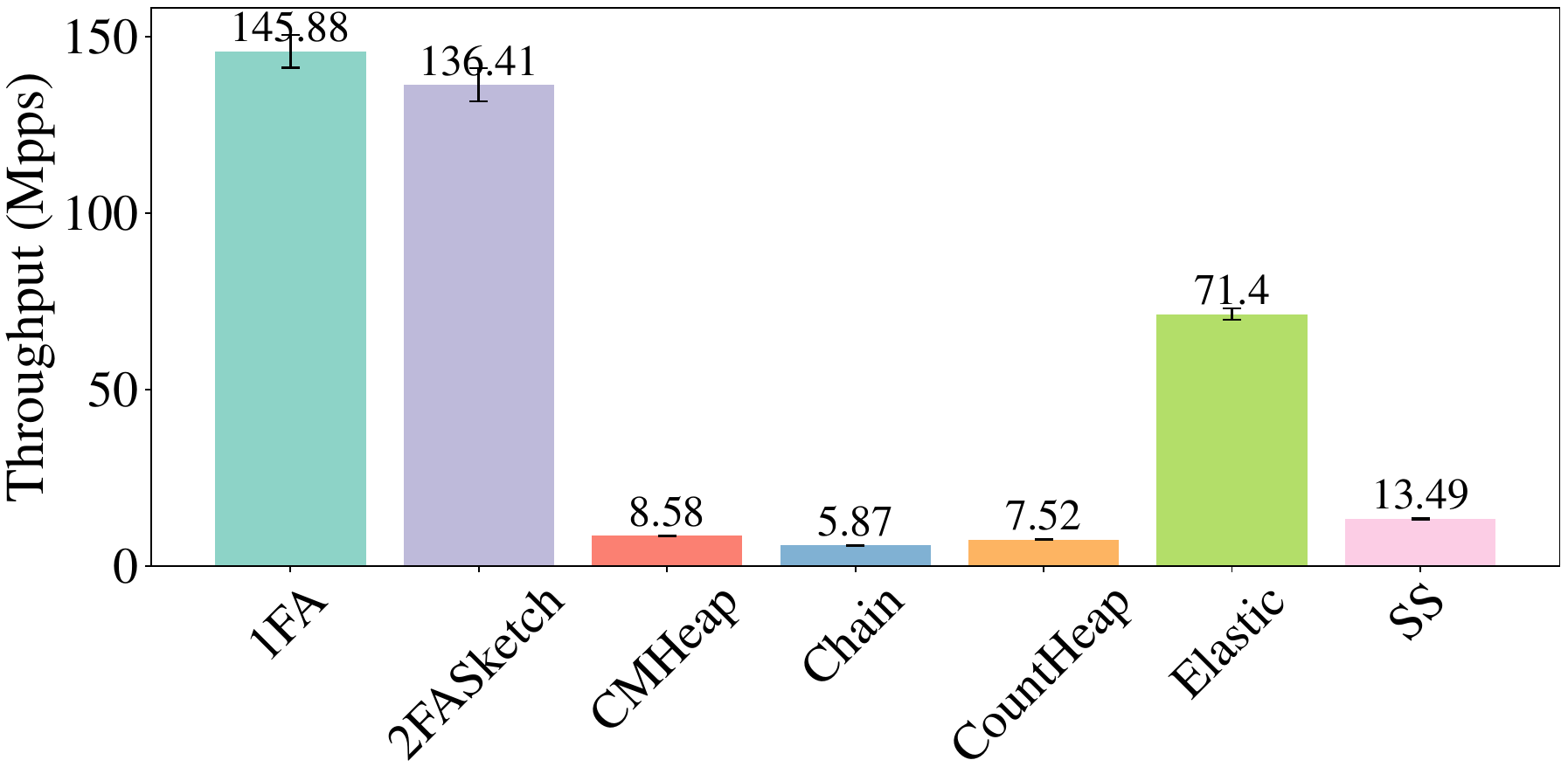}
    \caption{Throughput Comparison on CAIDA Dataset, including 1FA}
    \label{fig:throughput}
  \end{figure}

\presec
\bbb{Throughput (Figure~\ref{fig:throughput}):} Our experiments show that 2FA Sketch significantly outperforms five other algorithms on our CPU platform. It achieves nearly 140 Mpps with SIMD optimization, about 1.03 times faster than Elastic and over 9 times faster than conventional algorithms limited to 15 Mpps. Notably, Armor 2's rehashing, triggered only during intense conflicts, reduces throughput by just $6.5\%$ compared to 1FA.
\presec
\subsection{Evaluation on Parameter Settings}\label{Exp:parameter}
\postsec

\begin{figure*}[hbtp]

\centering
\begin{subfigure}[t]{0.24\textwidth}
\includegraphics[width=\linewidth, height=63pt]{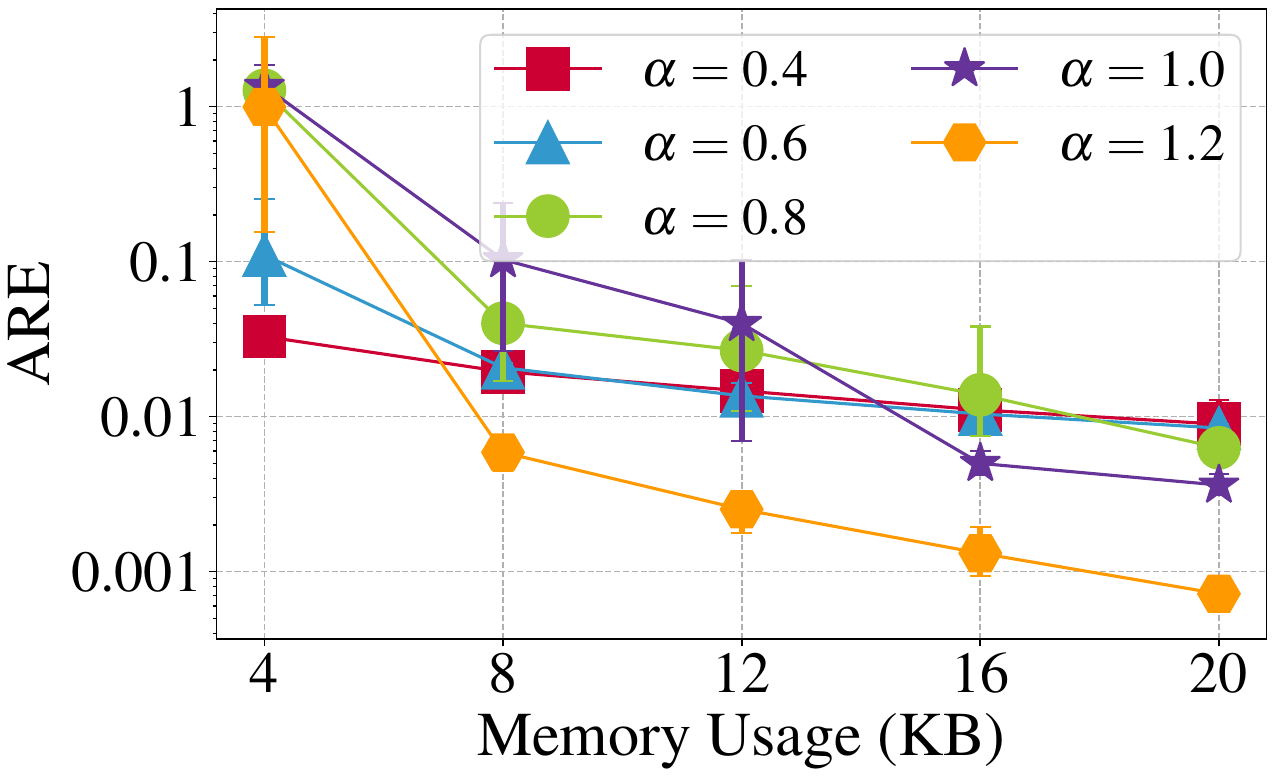}
\caption{ARE}\label{EXP:ZIPF_ARE}
\end{subfigure}\hfill
\begin{subfigure}[t]{0.24\textwidth}
\centering
\includegraphics[width=\linewidth, height = 63pt]{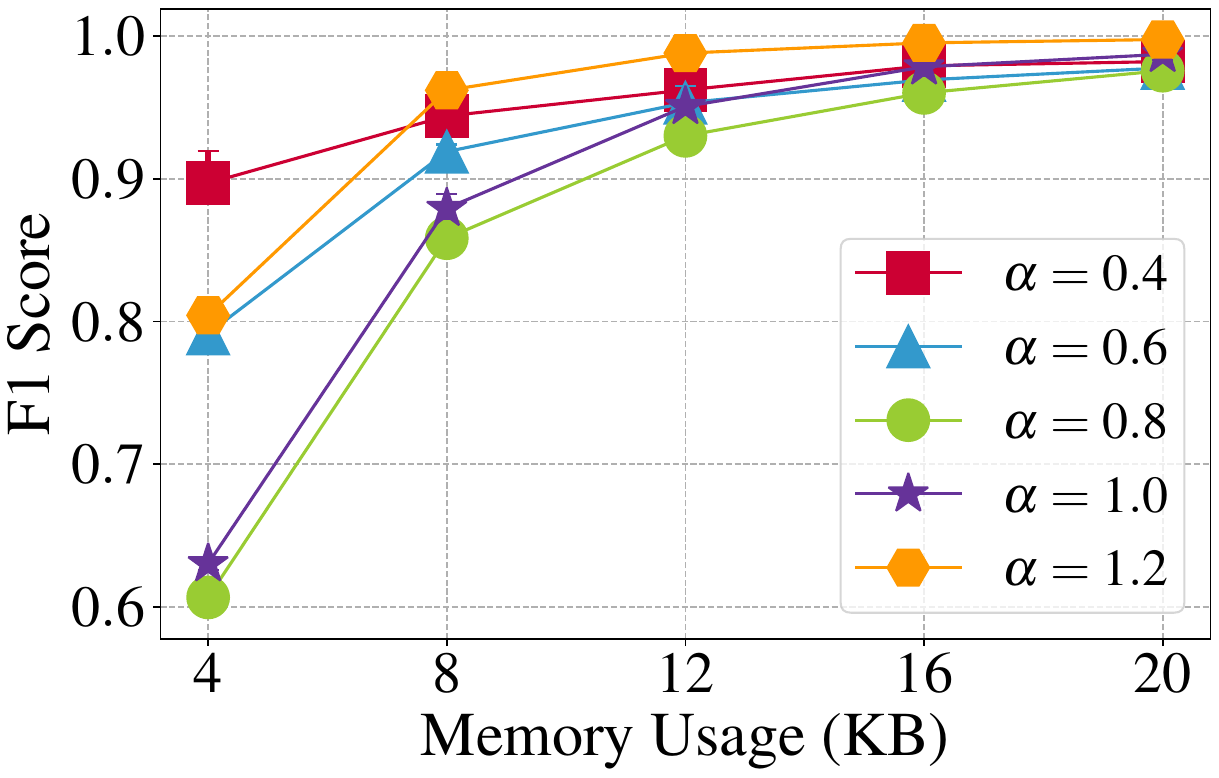}
\caption{F1 Score} \label{EXP:ZIPF_FS}
\end{subfigure}\hfill
\begin{subfigure}[t]{0.24\textwidth}
\centering
\includegraphics[width=\linewidth, height = 63pt]{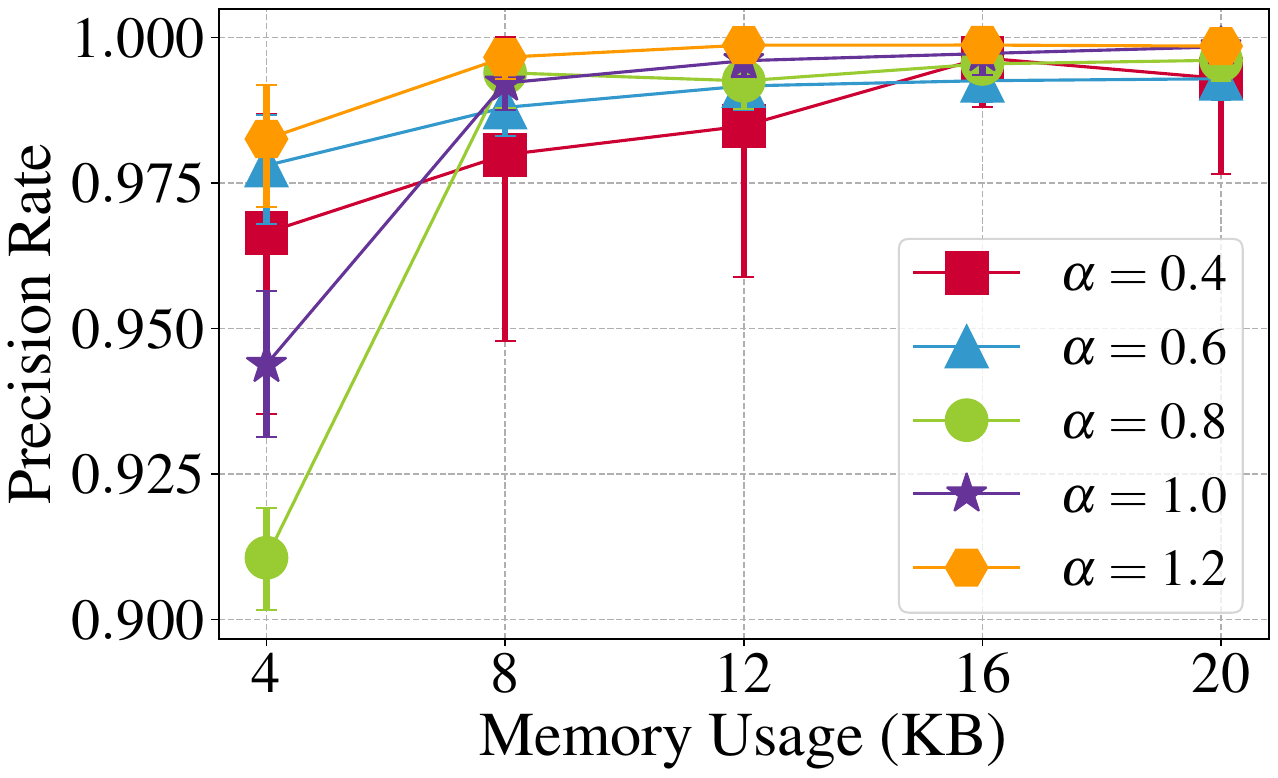}
\caption{precision rate}\label{EXP:ZIPF_precision}
\end{subfigure}\hfill
\begin{subfigure}[t]{0.24\textwidth}
\centering
\includegraphics[width=\linewidth, height = 63pt]{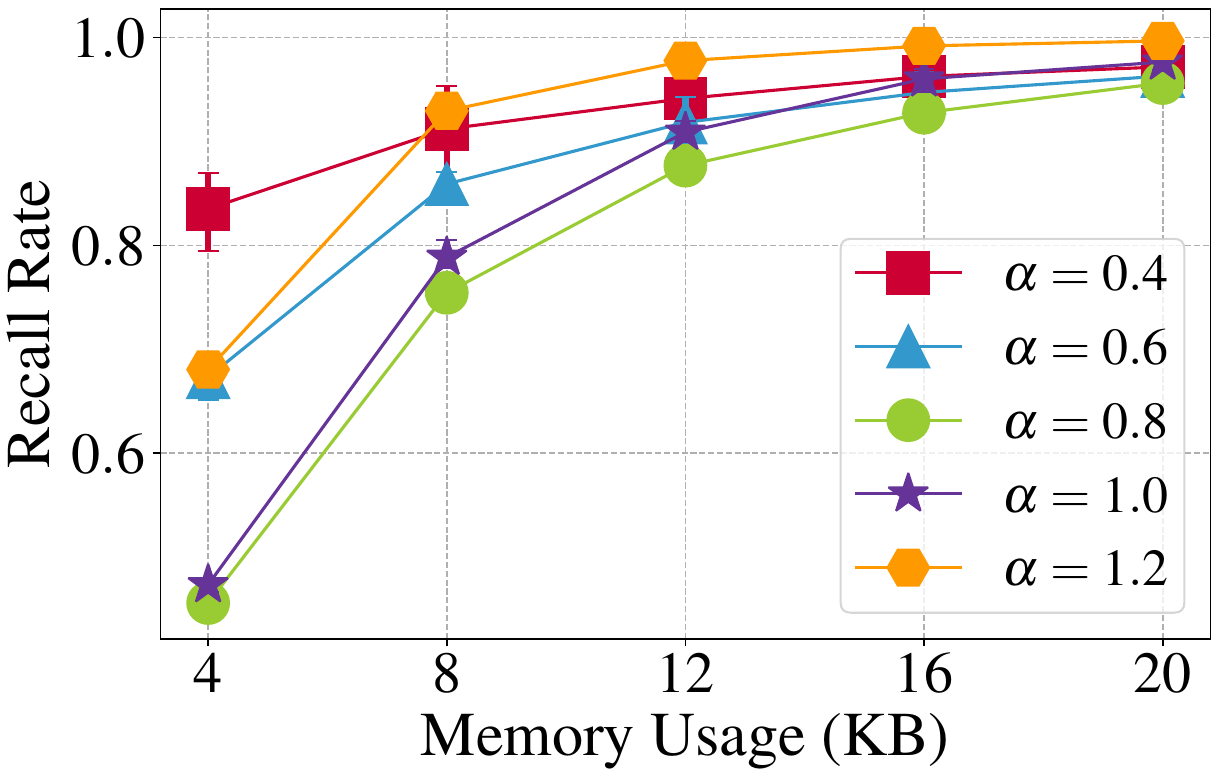}
\caption{recall rate}\label{EXP:ZIPF_recall}
\end{subfigure}\hfill
   \caption{Performance Analysis of 2FA Sketch with Varying Skewness $\alpha$ on  Synthetic Dataset}
    \label{fig:accracy_ZIPF}
   
    \vspace{-0.25in}
\end{figure*}

\bbb{Accuracy evaluation on 2FA Sketch with different skewness $\alpha$ (Figure~\ref{EXP:ZIPF_ARE}-\ref{EXP:ZIPF_recall}):} Our results demonstrate that 2FA Sketch maintains an F1 Score above $85\%$ with only 8KB of memory, but exhibits reduced performance at 4KB. Data streams with moderate skewness prove more challenging, characterized by a high number of flows and intense competition between large and small flows. 

\begin{figure*}[hbtp]
\begin{subfigure}[t]{0.22\textwidth}
\includegraphics[width=\linewidth, height = 66pt]{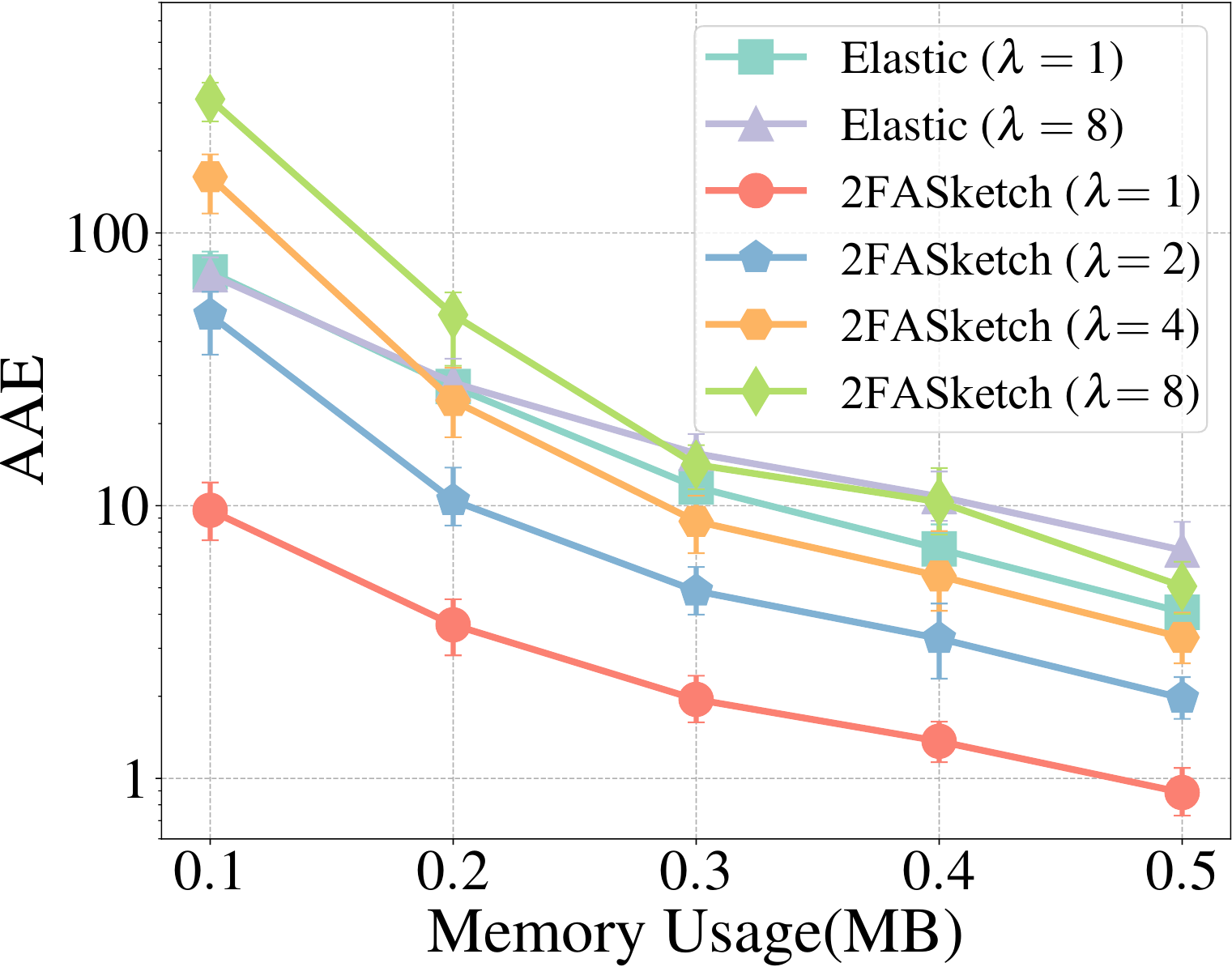}
\caption{AAE}\label{EXP:lambdaAAE}
\end{subfigure}\hfill
\begin{subfigure}[t]{0.22\textwidth}
\includegraphics[width=\linewidth, height = 66pt]{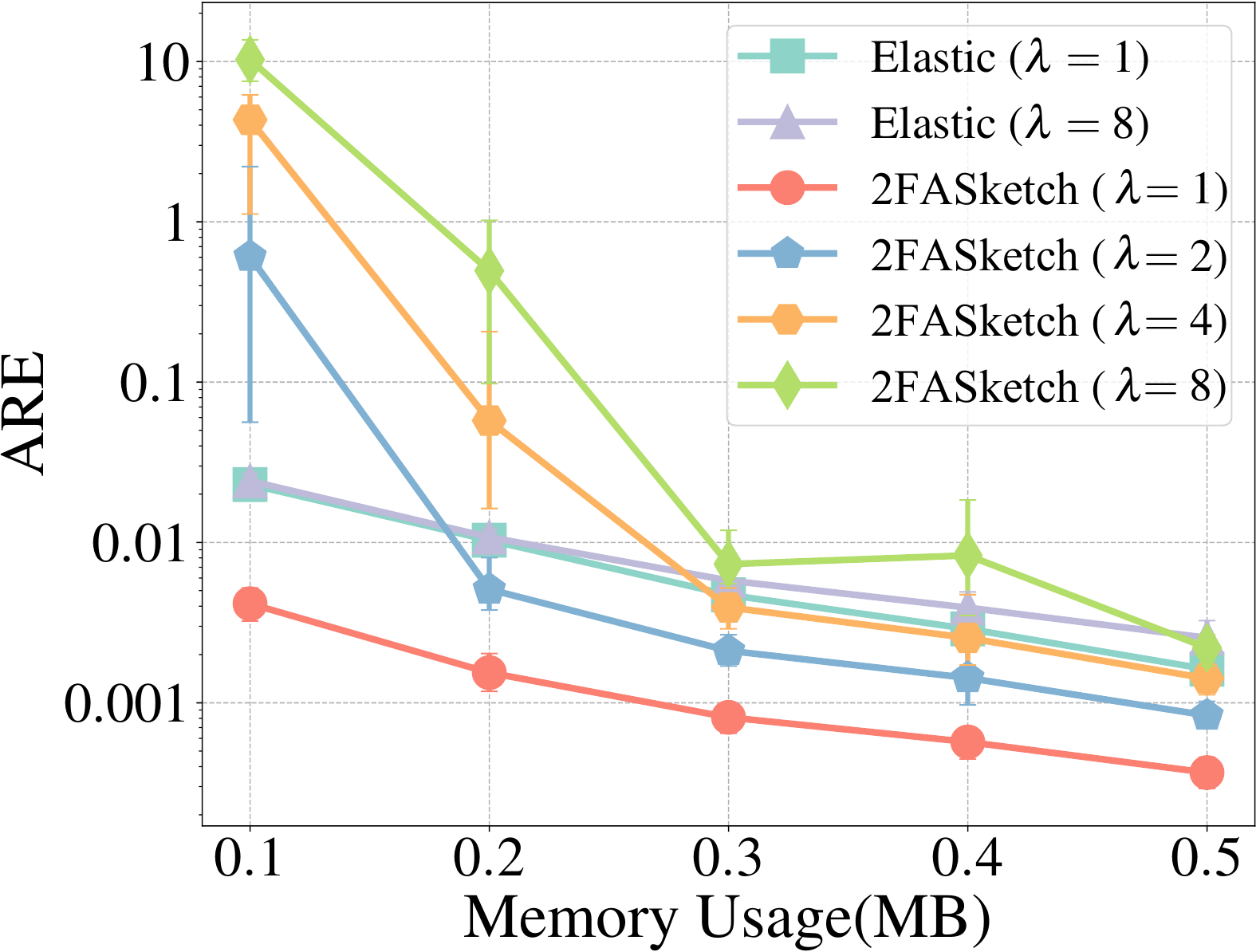}
\caption{ARE}\label{EXP:lambdaARE}
\end{subfigure}
\begin{subfigure}[t]{0.26\textwidth}
\includegraphics[width=\linewidth, height = 66pt]{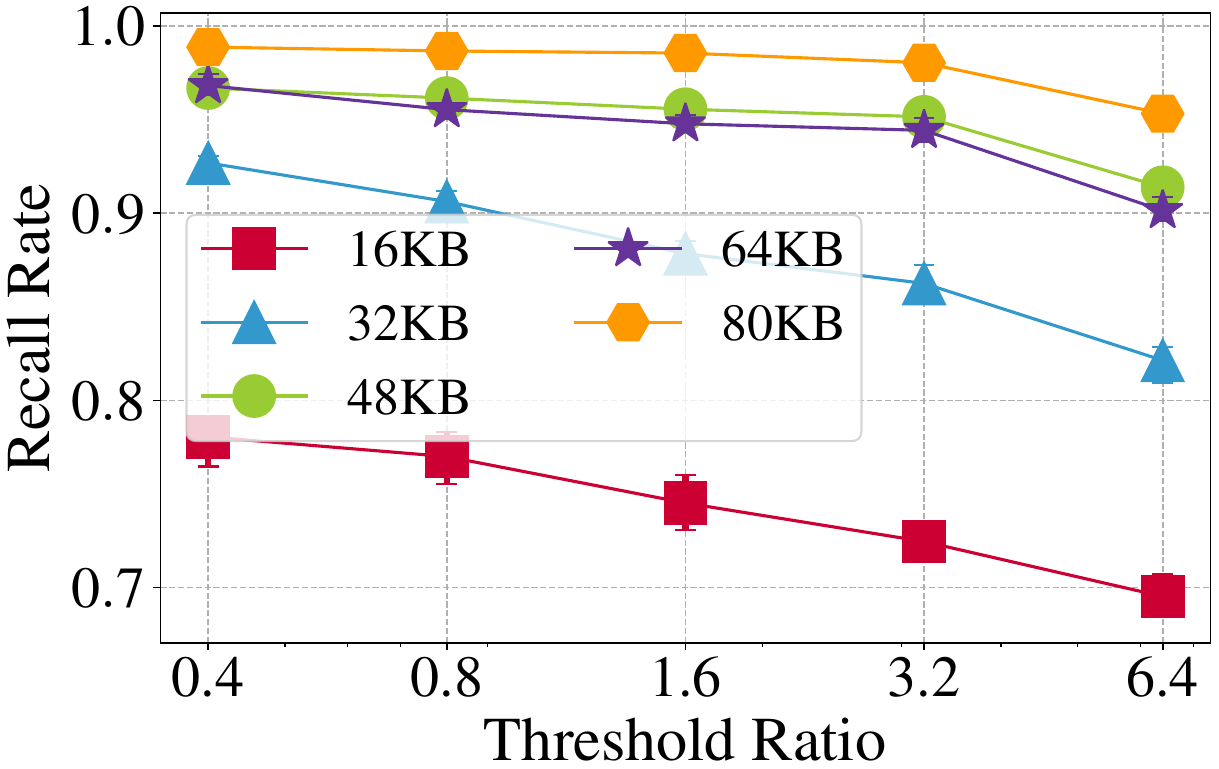}
\caption{Threshold Ratio vs. Recall Rate}\label{EXP:Theta}
\end{subfigure}
\begin{subfigure}[t]{0.26\textwidth}
\includegraphics[width=\linewidth, height = 66pt]{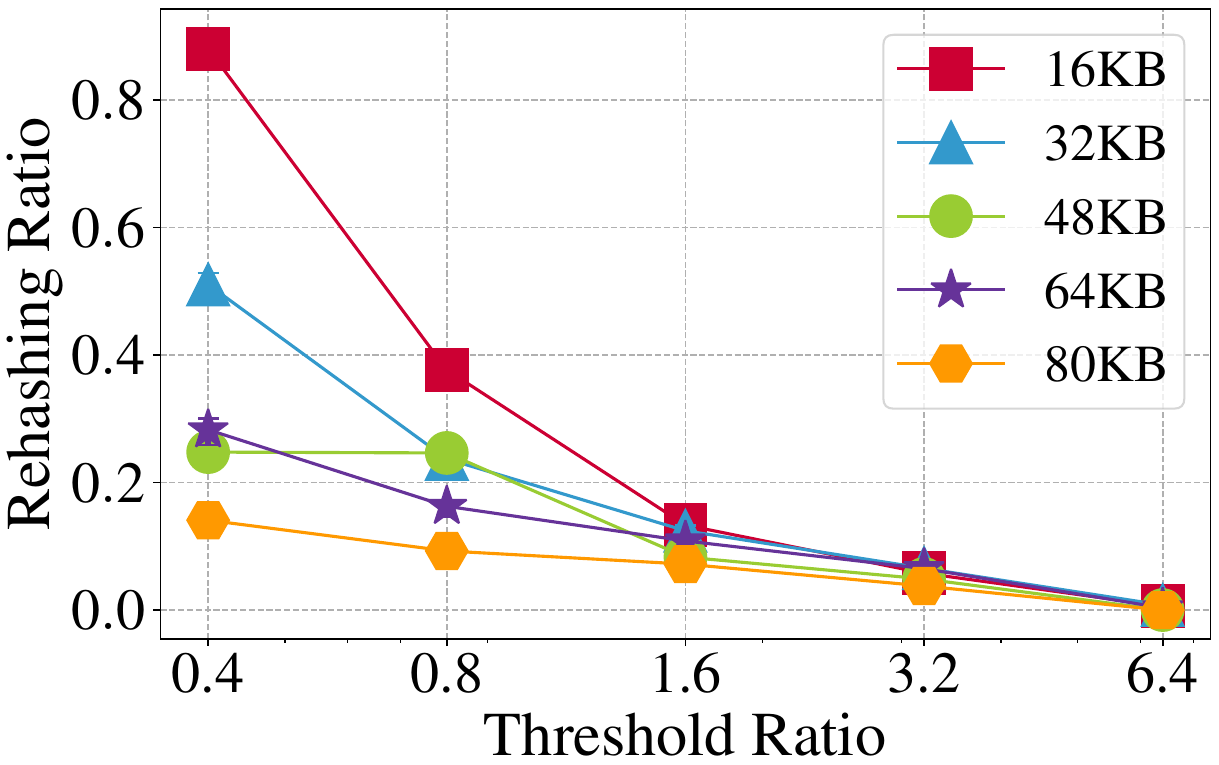}
\caption{Threshold Ratio vs. Rehashing Ratio}\label{EXP:Theta_ratio}
\end{subfigure}
   \caption{Performance Analysis of 2FA Sketch with Varying $\lambda$ and $\frac{\Theta_0}{\Theta}$ Ratios on CAIDA Dataset}
    \label{fig:accracy lambda}
    \vspace{-0.1in}
\end{figure*}

\bbb{AAE and ARE comparison of 2FA Sketch with different $\lambda$ (Figure~\ref{EXP:lambdaAAE}-\ref{EXP:lambdaARE}):} Our experimental results show that 2FA Sketch achieves highest accuracy at $\lambda=1$, with performance improving as $\lambda$ decreases. While Elastic outperforms 2FA Sketch at $\lambda=8$, it fails to improve at $\lambda=1$, where 2FA Sketch excels. At $\lambda=1$, 2FA Sketch achieves $4.4\sim8.2$ times smaller AAE and $4.4\sim6.9$ times smaller ARE compared to Elastic. These experiments were conducted with memory sizes ranging from 100KB to 500KB.

\bbb{RR evaluation with different threshold ratios(Figure~\ref{EXP:Theta}-\ref{EXP:Theta_ratio}):} Given Armor 2's significant improvement in RR shown in Figure~\ref{EXP:SM_recall}, we assessed the impact of varying $\frac{\Theta_0}{\Theta}$ on RR. Optimal RR is achieved with threshold ratios between 0.4 and 0.8, with a more pronounced effect under memory constraints. Considering the threshold ratio's influence on throughput, we also evaluated the rehashing ratio, observing a substantial decrease within the 0.4-0.8 range.


	\presec
\section{Conclusion}
\label{sec:conclusion}
\postsec

This paper introduced the Two-Factor Armor (2FA) Sketch, a novel data structure designed to enhance heavy hitter detection in high-throughput network environments. By addressing key limitations in existing Comparative Counter Voting approaches, 2FA Sketch provides dual-layer protection for improved accuracy and efficiency. The first layer, an improved Arbitration strategy, optimizes in-bucket competition through a theoretically derived optimal $\lambda$ parameter and a redesigned $vote^+_{new}$ as a conflict indicator. The second layer, a cross-bucket conflict Avoidance hashing scheme, further enhances heavy hitter detection by mitigating recall rate degradation caused by flow congestion. Experimental results demonstrate the significant performance improvements of 2FA Sketch over the standard Elastic Sketch, with error rates reduced by 2.5 to 19.7 times and processing speed increased by 1.03 times.

	
	\bibliographystyle{splncs04}
	\bibliography{InputFiles/reference}
\end{document}